# Machine Learning for the Physics of Climate

Annalisa Bracco[1], Julien Brajart[2], Henk A. Dijkstra[3], Pedram Hassanzadeh[4], Christian Lessing[5] and Claire Monteleoni[6,7]

[1]School of Earth and Atmospheric Sciences, Georgia Institute of Technology, Atlanta, GA, USA
[2]Nansen Environmental and Remote Sensing Center (NERSC), Bergen, Norway
[3]Institute for Marine and Atmospheric research, Utrecht University, Utrecht, the Netherlands
[4]Department of Geophysical Sciences and Commitee on Computational and Applied Mathematics, University of Chicago, IL, USA
[5]European Centre for Medium Range Weather Forecasts (ECMWF), Reading, UK
[6]INRIA Paris, France
[7]Computer Science Department, University of Colorado Boulder, Boulder, CO, USA
*e-mail:abracco@gatech.edu

**ABSTRACT**

An exponential growth in computing power, which has brought more sophisticated and higher resolution simulations of the climate system, and an exponential increase in observations since the first weather satellite was put in orbit, are revolutionizing climate science. *Big data* and associated algorithms, coalesced under the field of Machine Learning (ML), offer the opportunity to study the physics of the climate system in ways, and with an amount of detail, infeasible few years ago. The inference provided by ML has allowed to ask causal questions and improve prediction skills beyond classical barriers. Furthermore, when paired with modeling experiments or robust research in model parameterizations, ML is accelerating computations, increasing accuracy and allowing for generating very large ensembles at a fraction of the cost.
In light of the urgency imposed by climate change and the rapidly growing role of ML, we review its broader accomplishments in climate physics. Decades long standing problems in observational data reconstruction, representation of sub-grid scale phenomena and climate (and weather) prediction are being tackled with new and justified optimism. Ultimately, this review aims at providing a perspective on the benefits and major challenges of exploiting ML in studying complex systems.

**Key points:**

- *The use of Machine Learning is poised to transform the climate physics field.*
- *Major advances so far have occurred in extending observational data records in time, space and observables.*
- *Innovative approaches in sub-grid scale parameterizations may soon contribute to new (hybrid) climate models.*
- *Classical predictability barriers have been broken.*
- *Weather forecasting skills have improved, at a fraction of computing resources.*

**Website summary:** With the availability of big data and increasing computational power, methods from artificial intelligence, specifically machine learning, are being massively applied to climate physics. We focus here on novel results obtained so far in reconstruction, sub-grid scale parameterization and weather/climate prediction, and remaining challenges.

## Plain Language Summary

This review article covers the broader accomplishments of Machine Learning (ML) in the climate physics realm, and provides a perspective on the benefits and major challenges of exploiting ML advances. The intent is for both the limitations and opportunities highlighted to be relevant to other areas of physics broadly, and fluid dynamics more specifically.

## 1 Introduction

The climate of our planet, usually defined as the average weather over a period of years, constrains the weather we get. Accurate predictions of the climate system trajectory are a crucial science priority of the coming decades. Society needs detailed regional projections of future weather and climate extremes to better inform mitigation and adaptation strategies, and constrained estimates of the likelihood of reaching climate tipping points; assessments of impact and feedbacks of natural and engineered solutions to the climate challenge; and estimates of the uncertainties, risks, and economic and social impacts associated to deep cuts in emissions and the use of carbon removal technologies at scale[1,2]. Observing and modeling the evolution of the climate system, or any complex system, however, are hard tasks. The climate system is *multi-scale*, i.e., involves nonlinear processes characterized by spatial and temporal scales that differ by many orders of magnitude, and *high-dimensional*, i.e., involves many degrees of freedom that are coupled to each other. Machine learning (ML) has been rapidly advancing these tasks through applications centered around three themes: extending or better interpreting observations, advancing the development of parameterizations of small-scale processes (e.g., turbulent motions), and accelerating or improving multi-scale predictions. A schematic of these three pillars, and major research topics where ML has brought significant improvements, even major breakthroughs, in the past five years are introduced in Figure 1.

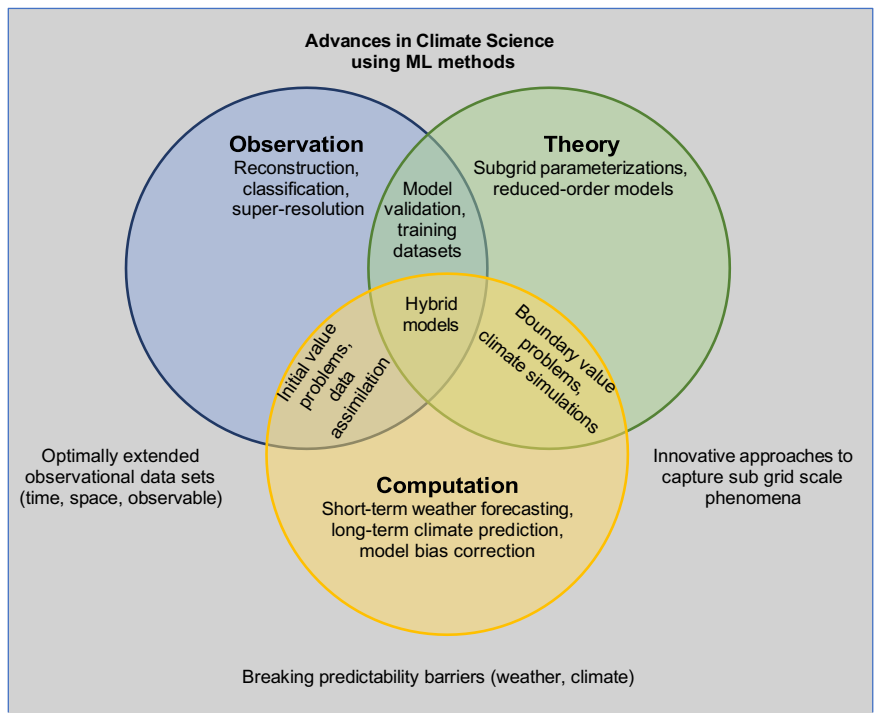

**Figure 1.** Advances in climate science linked to ML applications can be viewed as pertaining to three essential general areas and their intersections: observation, theory, and computation.

The central challenge of climate science lies in the climate system complexity. Climate prediction is a boundary value problem. In the climate case, the biggest uncertainties lie in determining how the energy received from the Sun (the boundary value) is distributed across the system's components, i.e., the atmosphere, ocean, land and cryosphere. Processes within and between each component, all the way to the ecosystems that populate them, interact and feedback on each other, modifying how the energy received is radiated back, absorbed or emitted. For example, the transfer of mechanical energy in the ocean, from the large scales of boundary forcing to the small scales of molecular dissipation, results from nonlinear interactions between mesoscale turbulence, submesoscale vortices, filaments and frontal structures, and gravity waves[3]. All these physical processes contribute to the drawdown of carbon from the atmosphere into the ocean, directly and indirectly through the contribution of the marine ecosystem, and in turn modulate how much heat-trapping greenhouse gas molecules are left in the atmosphere. In the atmosphere, a comparable problem is found in the interactions between clouds and the global atmospheric circulation. At a microphysical level, the early stages of cloud formation exhibit the properties of a colloidal system and depends on the atmospheric chemical composition, which varies with changes in anthropogenic emissions. Different types of clouds and clouds with different chemical compositions reflect solar energy differently, impacting the overall circulation, and clouds feedback on the climate system influencing the amplitude and spatial patterns of global warming. Boundary value problems are common to many physical fields, from fluid dynamics, to quantum mechanics, astrophysics and electromagnetism, and to many engineering disciplines.

ML has opened new avenues to address the climate prediction challenge, with enormous activity over the last decade dispersed over many different areas, from applied mathematics, computer science, and climate physics or climate modeling. Much of this activity has targeted specialized research communities[4–14]. Here, we review major advances enabled by ML, together with a perspective on the benefits and challenges it offers. We focus in the climate physics realm, where insuring generalization and interpretability, representing how patterns populating turbulent flows feedback on the large scale circulation, and preserving symmetries, conservation laws and physically balanced states when they exist, are critical requirements. While climate physicists and fluid dynamicists are our targeted audience, we trust limitations and opportunities highlighted to be relevant to other areas of physics, as well and of interest to nonlinear and computational physicists more broadly.

## 2 Application Themes

### 2.1 Machine Learning for Data Reconstruction, Downscaling, and Assimilation

The observation of the climate system is key to understanding climate physics and modeling climate change, and it is achieved by monitoring essential climate variables[15], such as temperature or precipitation. Similarly to lab experiments, observational data are, however, often incomplete in space, time, or scale, raising the need for "reconstruction". Missing value problems are common in all domains that deal with data and are a standard application of ML tools, following the successes in image reconstruction. The basic idea is to apply ML to transform incomplete data into corresponding complete sets. Among the many successes implied by the use of ML are the ability to relax constraints of traditional reconstruction methods such as linearity and Gaussianity; the access to effective generation techniques that allow for producing large ensembles of realizations for a fraction of the computational cost; and the possibility to derive new physical quantities not directly observed and, therefore, poorly modeled from available observations. A major limitation of most ML applications for reconstruction is that the outcome is usually constrained by observations alone, and there is no guarantee that physical symmetries or conservation laws are fulfilled. This problem can be addressed by incorporating physical constraints into neural networks [16,17] (see also Methods). Inferring the underlying laws based on measurement data alone, however, poses a crucial challenge: such reconstructions often involve ill-posed transformations and can be subject to numerical artifacts [18]. On the other hand, enforcing physical constraints (for example, preserving global precipitation sums, or ensuring monotonicity in given quantities, or guaranteeing mass conservation) improves the generalizability of models to future climate scenarios unseen during training, and their interpretability. It should be noted that while a definitive knowledge and understanding of symmetries and conservation laws is available for canonical turbulent flows, uncertainties and limitations stemming from the choice of physics-based model structure still hamper climate science.

---

**Box 1: Improving the best linear unbiased estimator (BLUE) with machine learning**

In general terms, in a reconstruction problem, we are estimating a vector **x**, which represents the state of a system. We have an imperfect knowledge of the vector through a "background" term

$$\mathbf{x} = \mathbf{x}_b + \mathbf{\nu}, \qquad (1)$$

where **ν** follows an normal law $\mathbf{\nu} \sim N(\mathbf{0}, \mathbf{C}_{bb})$. We assume we have access to partial observations (vector **y**) that is linked to the state **x** with

$$\mathbf{y} = \mathbf{H}\mathbf{x} + \mathbf{\varepsilon}, \qquad (2)$$

where **H** is a linear observation operator (e.g. subsampling) and noise **ε** follows a normal law $\mathbf{\varepsilon} \sim N(\mathbf{0}, \mathbf{C}_{yy})$. We are interested in the maximum a posterior $p(\mathbf{x}|\mathbf{y})$ probability which has a mean of $\mathbf{x}_a$ and a variance-covariance denoted $\mathbf{C}_{aa}$. By deriving the Bayes' formula and maximizing the probability[19], we obtain the best linear unbiased estimator (BLUE):

$$\mathbf{K} = \mathbf{C}_{bb}\mathbf{H}^T (\mathbf{H}\mathbf{C}_{bb}\mathbf{H}^T + \mathbf{C}_{yy})^{-1} \qquad (3)$$

$$\mathbf{x}_a = \mathbf{x}_b + \mathbf{K}(\mathbf{x}_b - \mathbf{H}\mathbf{x}), \qquad (4)$$

$$\mathbf{C}_{aa} = (\mathbf{I} - \mathbf{K}\mathbf{H})\mathbf{C}_{bb}, \qquad (5)$$

where **I** is the identity matrix, and **K** is known as "Kalman gain".

This formalism and its various implementations (Krigging, Kalman filter, Gaussian processes) have proved to be effective and scalable. Nevertheless, it relies on strong assumptions (e.g., linearity, Guaussianity, known error statistics) that can be relaxed with ML to extend the application of reconstruction problems. For example, generative models can be used in a Bayesian framework[20], and recurrent neural networks have been applied for spatial interpolation[21](Fig. 2A). Additionally, ML has been employed to mitigate the computational requirements of traditional data assimilation methods [22] (see also section 2.3).

### 2.1.1 Reconstruction in space (spatial interpolation)

Following the launch of the U.S. NASA Earth observation satellite "Landsat-1" in 1972, remote sensing satellites have revolutionized our ability to observe the climate system. Satellite data, however, offer only partial coverage in space. For example, sea surface height can be observed along the track of the satellite (Fig. 2A) or with uniform coverage but for limited satellite swaths. Sea surface temperature, on the other hand, can be measured through infrared (IR) sensors with good spatial (1-4 km) and temporal (10-15 min) resolution but only in cloud-free areas, or with microwave sensors, which may see through clouds, but have resolution of ≈ 20 − 50 km and cannot be used in coastal areas. Partial structural coverage in space and/or time and limited resolution at the boundaries hinder many other fields, from observational cosmology to molecular cell biology. In the climate realm, reconstructions of missing data are needed not only for monitoring purposes but also for computing global statistics and budget analyses (e.g. momentum, heat and mass), which are essential for boundary value problem systems[24]. Traditionally, reconstructions are done through spatial interpolation or data assimilation [19] as illustrated in Box 1. ML has contributed through algorithms originally developed for *inpainting* in computer vision, which solve the task of reconstructing missing regions in an image. Examples of applications of *inpainting* algorithms in climate include the reconstruction of biogeochemical or physical variables from sparse observations[21,25–27] that outperform traditional algorithms (see Box 1) such as Krigging.

### 2.1.2 Reconstruction in time (temporal interpolation)

Many datasets collected for climate monitoring lack time continuity, as is often the case in laboratory experiments. This may be due to limited sampling capabilities, sensor deficiencies, or intermittent instrument unavailability. For example, Argo floats (https://argo.ucsd.edu/) have revolutionized how we observe the ocean but profile the water column every ten days instead of continuously; climate model outputs are generally stored as monthly averages, instead of daily or hourly, due to storage costs; and sea ice thickness (Fig.2b) can be measured by combining data from two satellites, but only in winter. Similar problems exist in planetary science, where data collected by space probes share the limitations of satellite-derived observations[28].

In many cases, time interpolation can be achieved using a numerical model. If no accurate model is available or the interruption in data collection is too long or far back in the past, observed proxy variables can be used to infer the unobserved variable. This application resembles what was done for videos in the past, and ML allows to quickly evaluate several combinations of input proxies. Successful examples include extending the archive of phytoplankton[29] or sea ice thickness[30] to uncovered periods of the year or back in time. Optical flow methods, currently preferred for image processing, are instead suboptimal in many climate applications[31]. This is because the cutoff between temporal and spatial interpolation in geospatial data may be fuzzy and edges less pronounced than in videos.

### 2.1.3 Reconstruction of scales (downscaling, superresolution)

Downscaling refers to the disaggregation of coarse resolution data with the help of mathematical tools to infer high-resolution information. For instance, sea ice thickness (Fig.2C) is observed at an effective resolution of about 100 km[32], but important features, such as leads and ridges, occur at much finer scales. Storms remain poorly resolved in climate models, but downscaling the evolution of their statistics into the future is essential for developing mitigation and adaptation strategies. Traditionally, downscaling is realized through dynamical modeling: a (regional or local) high-resolution numerical model is run forced by boundary conditions provided by a coarser-resolution model or from an observational product. Dynamical downscaling has improved and will continue to refine our understanding (e.g.[33]), but is computationally expensive and limited in the time and space resolution that can be simultaneously achieved[34]. ML offers an effective complement through techniques often referred to as "superresolution" methods, following the computer vision convention. Recent ML downscaling applications pertain to precipitation[35,36], clouds[37], wind[38,39], solar irradiance[40], temperature in the atmosphere[39,41], and to surface data including sea surface height[42,43] in the ocean. Most ML algorithms require high-resolution and low-resolution fields paired in time for training, but probabilistic *domain alignment* has removed such need, as long as both fields are available[44,45]. More recently, superresolution techniques have been coupled with physical constraints to ensure, for example, energy or mass conservation across the low and high-resolution realizations [46].

### 2.1.4 Reconstruction as a probabilistic problem

Originally, many ML-based methods adopted in climate science did not address uncertainty quantification, which is a key aspect of traditional dynamical downscaling. Probabilistic ML and generative algorithms (see Methods section), however, allow to train ML models that sample ensembles from a target distribution and quantify uncertainty for a fraction of the cost of running climate model ensembles. Specifically, the reconstruction task can be posed as a probabilistic problem in which the objective is to train a neural network that approximates a conditional probability distribution $p(\mathbf{y}|\mathbf{x})$. Here $\mathbf{y}$ is the target quantity, e.g., a spatiotemporal field, and $\mathbf{x}$ are the known features; in the case of downscaling, $\mathbf{x}$ are the low-resolution fields. Often, the probability distribution $p(\mathbf{y}|\mathbf{x})$ is not known and cannot be determined, making the problem intractable a priori.

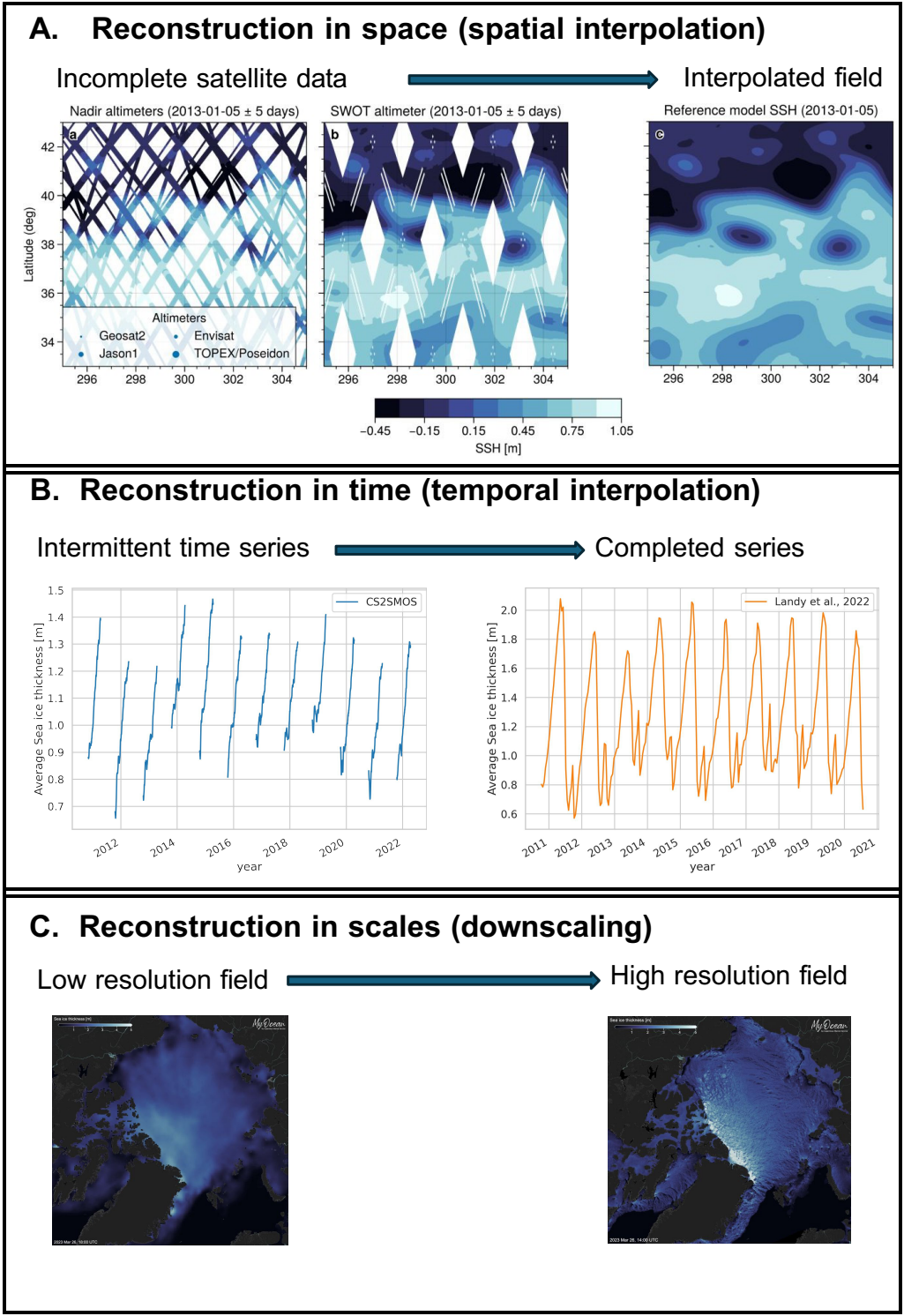

**Figure 2.** Examples of spatial, temporal, and scale reconstruction using ML. A. Simulated nadir altimeters (left) and SWOT sea surface height data (center) for a 10 days period based on a reference simulation (right) in the North Atlantic. Credits CLS/IGE. B. Time series of average Arctic Sea Ice thickness by the observational product CS2SMOS when no reconstruction is available in summer and with a ML-based reconstruction[23]. C. Map of sea ice thickness from a low-resolution observation product (CS2SMOS) (left) and a high-resolution model simulation (neXtSIM) (right). Source: https://marine.copernicus.eu/ Credit: CMEMS.

Variational inference techniques [47] can overcome the issue. Traditionally, it is assumed that the probability distribution follows a known function, generally a Gaussian, which allows for the application of methods like Kriging[48] or Gaussian processes[49] (see Box 1). In a ML regression framework, the problem reverts to estimating the moments of the distribution, e.g. to find the estimator $\mu$ of the expectation $E[\mathbf{y}/\mathbf{x}]$. The probabilistic reconstruction problem can be then simplified into the deterministic problem of computing $\mu$ as a function of $\mathbf{x}$. The drawback is that the estimated $\mu$ may be physically unrealistic, blurry, or smooth[48], as a direct consequence of computing an average or minimizing the least square, or the mean absolute error[50].

A smooth reconstruction, and the Gaussian distribution assumption, are unrealistic for a wide range of climate and, more generally, physics problems, such as predicting extreme or long-tail events, or forecasting the evolution of a chaotic system like weather [51]. A simple approach to sidestep the issue consists of extending the regression to estimate more moments of the distribution, such as variance or quantiles [50]. For a more general solution, generative ML allows to produce samples from the probability distribution $p(\mathbf{y}|\mathbf{x})$ as exemplified in [45] or the recent downscaling work by[41].

The use of probabilistic reconstruction has opened the opportunity to represent rare or extreme events. Conceptually, a generative algorithm could generate rare samples , as demonstrated in an idealized setting for a truncated Korteweg–de Vries system [52]. In practice, however, these approaches in more complex systems are limited by the training set if the extreme events are undersampled and their statistics may change over time. Strategies may be needed to augment the training set and enable a better representation of extremes, as in[53] (also see Current Challenges and Future Directions).

## 2.2 Data-driven subgrid-scale (SGS) Parameterizations

Climate science has advanced to the point that quantitative projections of the future evolution of the global climate are routinely provided on the base of climate model outputs, but those projections remain uncertain at a time when they are needed to support scientific understanding, political choices and economical decisions. The multiscale nature of the climate system requires many dynamical processes and interaction mechanisms to be properly represented. For example, ocean turbulent processes at kilometer scale impact the regional transfer of momentum and tracer properties[54], whereas convective clouds in the atmosphere occupy a size continuum from centimeters to hundreds of kilometers[55]. Due to incomplete understanding of processes occurring at very small scales and finite computational resources, many such processes (subgrid-scale or SGS, hereafter) are approximated or *parameterized* in terms of their impacts on the resolved scales. In many areas of science and engineering the development of SGS parameterizations (Box 2) is a fundamental problem since the advent of scientific computing, with earlier work done in the context of turbulent flows. SGS parameterizations in climate models have been traditionally derived from empirical relationships or idealized theoretical formulations. They interest not only the flow fields but also the tracers advected by them, from temperature and salinity to greenhouse gases and aerosols, and the interfaces of component of the Earth system (e.g., atmosphere-ocean-ice interactions). Parameterizations suffer from parametric and/or structural uncertainties. Machine learning is contributing to improve parameterizations by efficiently capturing the information contained in high-resolution simulations and/or observational data.

There are a number of different ways to use ML and data (be those observations or high-resolution model simulations) to improve the SGS parameterization. For example, if a current physics-based parameterization suffers from parametric uncertainty, then ML and data can be used to better estimate these parameters. On the other hand, if the parameterization suffers from structural uncertainty (e.g., incorrect equation), then ML can be used to learn the entire functional form of the relationship between the resolved and SGS processes from data (Box 2). Both of these procedures can be done *offline* or *online*. In the former, akin to supervised learning, the SGS parameterization is developed *decoupled* from the climate model and is coupled after training. The loss function is often the mismatch between the "true" and parameterized SGS terms (e.g., flux or forcing, ∏ in Box 2). In this section, we use the term "online learning" to denote methods in which the parameterization is developed while it is *coupled* to the climate model, and the loss function is usually the mismatch between the trajectories or some statistical measures of the true/desired climate and of the simulated climate (note that the standard definition of *online learning* as a family of ML algorithms will be used in subsequent sections). These approaches each have their own strengths and weaknesses, and their application is the subject of extensive ongoing investigations. Below, we discuss some key examples.

> **Box 2: SGS parameterization with machine learning:**
> Suppose a nonlinear system with state vector **x** is governed by the nonlinear partial differential equation (PDE)
>
> $$\frac{\partial \mathbf{x}}{\partial t} = F(\mathbf{x}). \tag{6}$$
>
> Coarse-graining of this equation with some kind of low-pass filter (denoted by the overline) leads to the following equation if we attempt to write the equation explicitly in terms of $\bar{\mathbf{x}}$
>
> $$\frac{\partial \bar{\mathbf{x}}}{\partial t} = F(\bar{\mathbf{x}}) + \underbrace{\overline{F(\mathbf{x})} - F(\bar{\mathbf{x}})}_{\Pi}, \tag{7}$$
>
> where $\Pi$ is the subgrid-scale (SGS) term (often written in terms of fluctuations, $\mathbf{x}' = \mathbf{x} - \bar{\mathbf{x}}$). The goal of a parameterization, or closure modeling, is to formulate $\Pi$ as an explicit function of $\bar{\mathbf{x}}$ such that Eq. (7) only depends on the large scales, which are computationally more tractable. One of the most common, and still used, class of parameterizations for momentum transfer in climate models is the eddy diffusivity class of closures (such as the Smagorinsky model), which assumes $\Pi = \nu \nabla^2 \bar{\mathbf{x}}$. Using this as an example, the goal of parameter optimization is to find the best $\nu$ that characterizes viscosity in terms of $\bar{\mathbf{x}}$ such that solutions of Eq. (6) match some key statistics of the solutions of Eq. (7); i.e., minimizing the loss $\|S(\mathbf{x}(t)) - S(\bar{\mathbf{x}}(t))\|$, where operator $S$ calculates some statistical properties of the solutions of Eq. (6) or Eq. (7) over a long period (e.g., mean, spectrum). Online learning of parameterization uses the same loss function but with $\Pi$ approximated by a neural network $N(\bar{\mathbf{x}}, \theta)$ with $\theta$ representing the weights. The goal of offline parameterization learning is to find a data-driven representation of $\Pi = N(\bar{\mathbf{x}}, \theta)$ without a priori assumptions about its structure by minimizing a loss function of the form $\|\Pi^{truth} - N(\bar{\mathbf{x}}, \theta)\|$, $\Pi^{truth}$ is obtained from the solutions of Eq. (6). Equation discovery often uses a similar loss function but aims to find a closed-form representation of $\Pi$ by learning the coefficients of terms in a library.

### 2.2.1 Parameter optimization

In most models representing complex systems, SGS parameterizations usually rely on a set of PDEs and often depend on poorly constrained parameters. In the climate modeling community, the process of estimating these parameters is known as tuning or calibration. One important ML application in climate physics is parameter optimization[56–59], by which the parameters used by each parameterization scheme are objectively tuned - offline or online - based on observations and/or high-resolution simulations, using, for example, a Gaussian process or ensemble Kalman inversion (EKI). This process replaces and optimizes the often subjective tuning based on the modelers knowledge about the feasible range through which each parameter may vary. Parameter tuning using ML requires, however, choosing an appropriate objective or loss function, for example global-average root-mean-square errors for several model variables, or Rossby wave propagation patterns[60], and the target to match (for example, rainfall, or cloud coverage statistics from an observational dataset), which together describe the optimization problem to be solved. In practice, there could be several parameterizations, each with several uncertain parameters, that have to be calibrated simultaneously. The objective function should aim at reducing systematic errors in the climate model while also preserving empirically observed relationships among variables and physically conserved quantities. Shortcomings are intrinsic to the assumption that only parametric uncertainty matters, and to the subjective choice of loss function and target. It is not uncommon to approach the target through error cancellation, especially if several parameterizations are tuned together. Practically, challenges emerge when estimating a large number of parameters: the problem is usually ill-posed and requires regularization, or the computational cost of the optimization algorithms is too high. Interpretability and generalization are also of concern, especially when the data available for training lie in a restricted region of the phase space, while the model may be used for predictions of other phase space regions, e.g., a warmer, unseen, climate[61].

### 2.2.2 Offline parameterization learning

The above optimization problem can be approached by learning the entire functional form of a parameterization with ML models. This is often done using deep neural networks (DNNs, see the Methods Section). In the offline framework, high-resolution simulations of the process to be parameterized - for example, atmospheric convection[62] or ocean turbulence[63] - are used to train and test the NN which is then coupled to the climate model. Data-driven parameterizations obtained in this way might have lower structural uncertainties compared to the physics-based parameterizations, with a low computational cost. Conservation laws and physical constraints can be imposed whenever unphysical behaviors emerge[16,64–66]. Lack of interpretability, need for long high-resolution simulations to accurately extract the SGS terms with enormous storage requirements, and the emergence of numerical instabilities after coupling to climate models are major limitations[63,67–70]. It is indeed common for DNNs to show much promise when evaluated offline, but limited performances when coupled online in realistic applications due to implementation issues, numerical instabilities, or error cascade (a small error at one step propagates across variables and grows larger across an entire process). Ongoing research is focusing on these problems. Attempts to partially address them for specific cases can be found in[66,71–74]. Most recently, a framework to automate the sampling and validation of coupled ML parameterizations has been proposed[75].

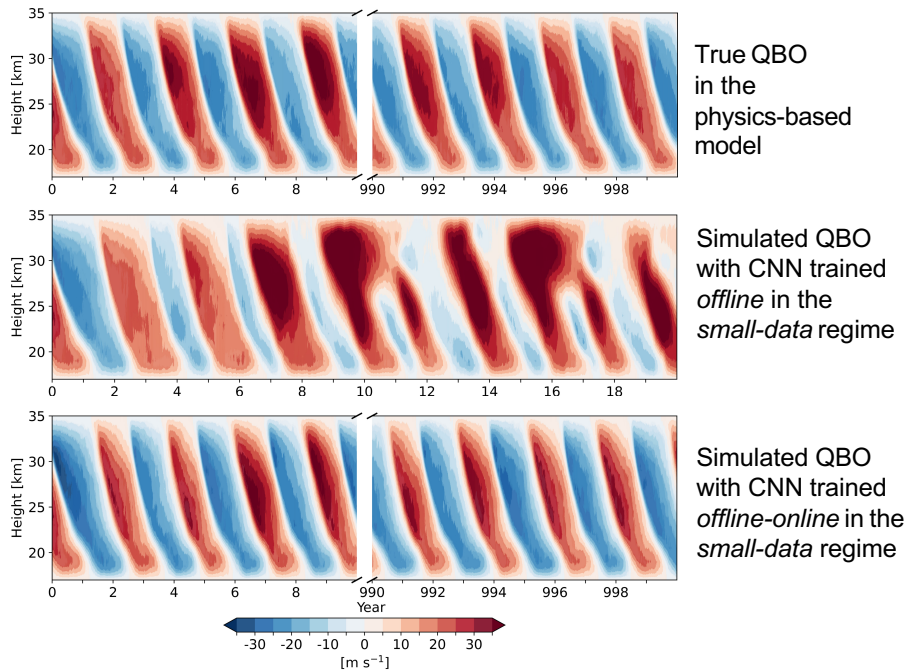

**Figure 3.** a) Height-year plot of zonal-mean tropical zonal wind showing the atmospheric Quasi-Biennial Oscillation (QBO) in a 1D stochastic model with physics-based parameterization of gravity waves. Replacing the physics-based parameterization with a convolutional neural network (CNN), see 3) trained offline in the big-data regime (100 years) yields very similar QBO statistics (not shown). b) CNN trained offline in the small-data regime (18 months) produces an unphysical and unstable QBO. c) The CNN from (b), once two of its layers are re-trained online, produces the correct QBO statistics. A parameterization learned from applying the common equation discovery approach also fails to produce the correct QBO unless a physics-informed library that accounts for the non-locality in gravity waves dynamics is employed (not shown). Plots are adapted from the results in[76]; courtesy of Hamid Pahlavan.

*2.2.3 Online parameterization learning*

The entire functional form of a parameterization can also be learned online, often using an over-parameterized DNN (a DNN with a number of trainable parameters larger than the number of training observations)[77]. This kind of architecture is usually preferred for its ability to generalize to noisy test data, and for its robustness[78]. In this case, detailed high-resolution simulations of the process of interest and the extraction of the SGS terms from such simulations are unnecessary, and only statistics from observations or high-resolution simulations are used. This approach aims at minimizing a loss function of the mismatch in statistics, and its greatest challenge is that the climate model has to be run for many times and/or for relatively long time while the ML model is trained. Additionally, instabilities may emerge during training or while solving the optimization problem. Recent papers have focused on using three methodologies: EKI[79], reinforcement learning, and differentiable modeling[77,80,81]. The new NeuralGCM developed by Google, for example, adopted differentiable modeling and end-to-end training and combines a conventional equation-based dynamical core with a NN that acts as a holistic parameterization predicting tendencies[74]. This hybrid GCM shows promise in reducing some of the biases of traditional GCMs, e.g., in frequency and trajectories of tropical cyclones. While advances in these methodologies may improve some of the optimization challenges, issues linked to error cancellation, the choices of targets and loss functions, and instabilities, remain.

All aformentioned approaches share a major weakness: they do not generalize out-of-distribution, i.e., they do not extrapolate to a system different from the training set. This is crucially important for climate change applications. The problem is most severe for online learning of a NN, because these extrapolate poorly[82,83], and less severe for parameter estimation (online or offline), especially if the variations in these parameters are small across climates. Offline learned NNs can partially address this issue by learning from a library of simulations that include climate change scenarios and incorporating physics constraints[65,69,84].

### 2.2.4 Mixed Approaches

For a given problem, the development of a ML parameterization is constrained by the level of theoretical understanding, the quality of the current physics-based parameterization, and the quality and quantity of the attainable high-resolution simulations and/or observational data, among other scientific and practical considerations. The approaches introduced so far are not mutually exclusive. For example, a DNN can be trained *offline* with a small amount of data from high-resolution simulations in an idealized or limited setting. The resulting data-driven parameterization may not be accurate enough for stable realistic simulations (see Fig. 3 for an example based on a 1D stochastic model of the Quasi-Bennial Oscillation (QBO), a gravity wave-driven mode of tropical stratospheric variability), but may still be useful. The NN can indeed provide reasonably good priors for online re-training with just observed statistics (which do not need detailed high-resolution simulations) using methods such as EKI , RL or differentiable modeling. The outcome of this offline-online learning sequence is a parameterization that yields stable and accurate simulations (see Fig. 3), while broadly addressing many of the shortcomings of the offline-only or online-only learning approaches[76].

### 2.2.5 Equation discovery

Equation discovery addresses the interpretability and extrapolation challenges and can be applied to observational data or high-resolution process simulations. With relevance to climate science, equation discovery has been used to find closed-form equations to parameterize ocean eddy momentum, temperature, and energy. Relevance vector machine (RVM), for example, has been used to identify the closure term in an idealized configuration of a primitive equation ocean model[63], and more recently a similar approach has been applied to cloud cover[85]. Recently, genetic expression programming and symbolic regression have been combine to identify a closure for quasi-geostrophic turbulence that performs as well as DNNs on the training domain but generalizes better, and that depends on higher-order derivatives of the mean velocity and potential vorticity[86]. Advantages of the equation-discovery approach compared to DNNs are the interpretability of the learned closures, potentially better generalization to other climates, efficient implementation in downstream computational models (given the sparse nature), better performances when noise affects the training data, and the need for smaller training sets [87]. The challenges of instability and worsening of performance in online versus offline applications, however, remain. This problem may be solved with opportune choices of the loss function that account for physical conservation laws in the system[70].

A different path, applied so far to idealized turbulent flows, consists in using ML to discover the whole mathematical description of the physical phenomenon investigated rather than a closure. The main idea behind is to use ML to identify parsimonious models in the form of coupled nonlinear PDEs from spatiotemporal data. For a weak formulation of differential equations, this framework uses sparse regression, and physical assumptions of smoothness, locality, and symmetry[88]. Regression alone becomes intractable in most cases, because the library of terms that can appear in a model grows exponentially with both the order of nonlinearity and the number of variables. Physical assumptions of smoothness, locality, and symmetry help constraining libraries[89,90]. More work is needed with this latest approach to extend it to climate-relevant applications.

## 2.3 Data-driven Prediction and Forecasting

Predicting the evolution of the climate system into the future is central to climate science[91]. As mentioned, climate prediction is a boundary value problem. Models must assume how the the conditions that constrain the climate evolution over the long-term will evolve, for example making educated guesses of future greenhouse gas emissions. Weather forecasting, instead, is an initial value problem, and depends first and foremost upon an accurate knowledge of the current state of the weather system. In both cases, uncertainty quantification is commonly achieved through ensemble runs that reduce the generalization error of the prediction/forecast by sampling from the distribution of possible trajectories (climate) or initial conditions (weather) of the system. Machine learning has brought new impetus to the field, with results beyond what was considered possible (Box 3). Advances have been especially rapid for short- and medium-range (∼ 14 days) weather forecasting, therefore included in this review, but there is a path in sight for extending these advances to climate time scales (annual, decadal and longer). Most recently, techniques from eXplainable AI (XAI), see Methods, have also provided insight into why particular ML models can increase forecasting skill[92].

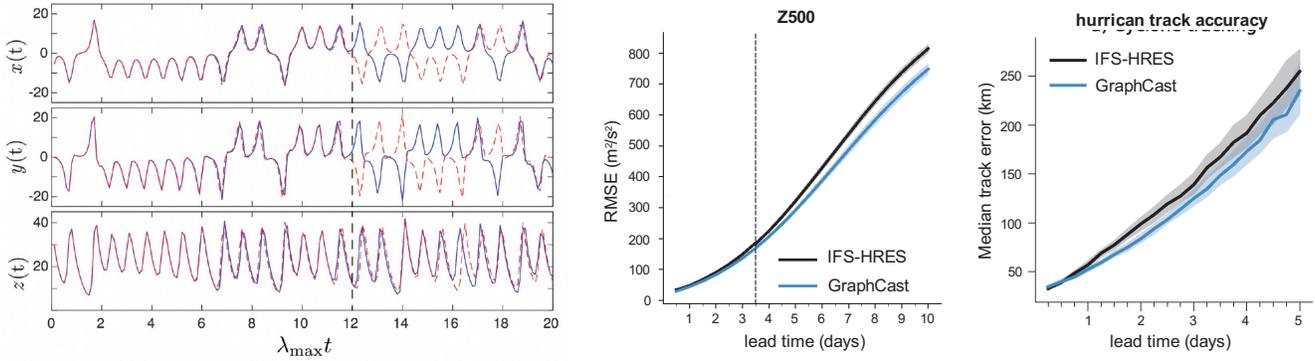

**Figure 4.** Left: Results[93] of a reservoir computer prediction (blue) and the 'true' trajectory (red) of the Lorenz system (Box 3); here $\lambda_{max}$ is the Lyapunov exponent. Right: Forecast skill of the GraphCast model[94] for geopotential height at 500hPa, which is often used to compare the skill of weather prediction models, and for hurricane positions, i.e. a form of extreme events that is rare in the training data. The European Centre for Medium Range Weather Forecasts (ECMWF)'s Integrated Forecasting System (IFS), an state-of-the-art physics-based model, is used as baseline.

---

**Box 3: Machine Learning in Forecasting**

The Lorenz-63 model has been used extensively as conceptual framework to reveal the essence of finite predictability within a chaotic system[91], and as a test case for both traditional and ML prediction techniques. It provides a simplified mathematical description of atmospheric convection and is based on three ordinary differential equations obtained through a Fourier series expansion retaining only the lowest order terms. It is given by

$$\frac{dx}{dt} = s(y-x), \qquad \frac{dy}{dt} = rx - y - xz, \qquad \frac{dz}{dt} = xy - bz$$

where the standard values of the parameters are $s = 10$, $r = 28$ and $b = 2.667$. Solutions quickly diverge for very small differences in the initial conditions. Mathematically the divergence rate of two trajectories is measured by the largest Lyapunov exponent $\lambda_{max}$, which, for the Lorenz-63 system, is positive and about 0.91. For lead times larger than the Lyapunov time $1/\lambda_{max}$, the skill of conventional methods typically quickly vanishes. Machine learning methods have shown that it was possible to build an emulator of a dynamical system with a predictive skill beyond the Lyapunov time, as exemplified in Fig. 4. In this case, training data are used to generate a dynamical system on a reservoir (with a network of 300 nodes) for $t < 0$. The reservoir is then treated as an autonomous dynamical system from $t = 0$ and produces a trajectory that remains near the 'true' Lorenz model trajectory for longer than the Lyapunov time.

---

### 2.3.1 Daily time scales: Machine learning for weather forecasting

One of the first papers to explore ML for medium-range weather forecasting[95] considered predicting the evolution of geopotential height without physical constraints. It was concluded that a data-driven, ML based global forecasting model, also known as emulator, while useful, would face limitations comparable to those of conventional models, e.g. with respect to the time step size, which was confirmed in subsequent works[96,97]. Better suited ML architectures improved both performance and efficiency[98,99], but did not reach the capabilities of state-of-the-art conventional models. Despite progress, skepticism about ML-based weather models persisted[100], especially regarding their ability to predict unseen weather states.

Several groups building ML emulators for medium-range forecasting[94,101–105], however, have now demonstrated that unseen weather states can be predicted from training on historical data, often with slightly better deterministic forecasting skill than the best conventional models for most variables and at low computational cost. The emulators perform well also for rainfall, compared to conventional models, but there remain deficiencies, not the least because high-quality training data are not available globally for this variable[106]. The ML models commonly employ the wind velocity components, temperature, and specific humidity on approximately 10 vertical levels as well as selected surface variables at 0.25° resolution, i.e. a much smaller state representation than conventional models. Important for their success is the training on the ERA5 reanalysis[107], which provides a long, consistent, high-quality dataset readily amenable to machine learning. The term reanalysis refers to the outcome of a variety of data synthesis often coupled to data assimilative models that incorporate observations from multiple sources and span an extended period of time. Improvements of ML compared to conventional models likely result from ERA5 being observationally better constrained than forecast models. A second crucial factor for the breakthrough resides in the use of larger and more sophisticated neural network architectures, e.g. Fourier neural operators[102], transformers[104], a combination of CNNs and transformers[103], and graph neural networks with a multi-resolution hidden mesh[94,101]. Two noteworthy recent examples include the hybrid NeuralGCM[74], already mentioned, and the local area model MetNet-3[108] that extends earlier work[109,110] on precipitation nowcasting (i.e., forecasting on a very short term period of up to 2 hours) and produces km-scale high resolution predictions up to 24 hours in the future. As input, MetNet-3 employs direct observations from ground stations as well as satellite data, partially side-stepping the need for conventional data assimilation.

Despite the rapid progress in ML-based weather prediction[111], open problems remain. For example in Pangu-Weather[103] the error growth of small amplitude oscillations is slower than observed[112], and the same model violates geostrophic balance and exhibits a significant discrepancy for vorticity and divergence fields compared to ERA5[113]. Group-symmetries for weather and climate exist in idealized systems, e.g.[114,115]. However, their pertinence to more realistic settings as in the ERA5 reanalysis, and consequently in neural networks trained on it, is unclear since ERA5 includes many types of processes that do not adhere to the group structure. Furthermore, only symmetries for low-dimensional Lie groups have been considered in the ML literature[116]. Some approaches[117,118] explored the use of symmetries in ML models and reported improved generalization. State-of-the-art models for weather, however, do not enforce them explicitly. The violation of conservation properties remains a hindrance in the extension of pure data-driven models to time scales of climate relevance. Research addressing these deficiencies is underway[119]. The surprising generalization capabilities of Pangu-Weather[120] suggest that the application of this kind of deep learning based models to climate may be possible. XAI techniques have been successfully applied to provide post-hoc explanations of the performance of ML models, for example for severe weather prediction[121], but the generalization of such methods for large-scale data-based emulators remains challenging[92].

### 2.3.2 Subseasonal to Seasonal Forecasting

Subseasonal to seasonal (S2S) prediction bridges the gap between the medium-range weather forecast and climate prediction (above one month). The time range offers societal application opportunities, from agriculture to water management and disaster preparedness. S2S forecasts are delivered less frequently - usually once a week - and for fewer variables than weather products.

One successful approach in the S2S context consists in continuously training a model as data becomes available[122]. Recently, for example, a computationally cheap weight determination was introduced[123] that allows for continuously adjusting probabilistic S2S forecasts in between production intervals using newly available data. Long known in the ML literature as *online learning*[124], this classification should not be confused with the term of the same name introduced in section 2.2.3. Online learning algorithms typically have far fewer parameters than deep learning models, requiring less computing power. They can be applied adaptively, combining climate model ensemble projections and observations when available, either at a global scale[125], or in a distributed fashion for ensemble predictions at a set of spatial locations[126,127]. The latter approach discovers and exploits relationships (in ensemble member skill) within spatial neighborhoods. For generating subseasonal forecasts at multiple lead times simultaneously, this neighborhood similarity can be extended to time by exploiting similarity between prediction tasks at lead times that are one month apart[128].

Data-based S2S forecasts of specific variables has already outperformed traditional models. Leveraging work on analog forecasting of extreme events[129], an extreme-focused CNN architecture was developed[130] for heat waves that predicts temperature anomalies for up to 28 days. The CNN emulator competes with physics-based forecast systems but is computationally much cheaper. Most recently, a global transformer neural network S2S model with a perturbation module for flow-dependent perturbations has outperformed the IFS forecast in key variables, including total precipitation and tropical cyclones[131]. Finally, with XAI methods it has been shown that neural networks can identify physically meaningful sources of S2S predictability[132].

### 2.3.3 Interannual scales: El Niño forecasting

At interannual scales, climate variations stem from the interactions among climate components. Often, these interactions result in preferred states in observable patterns, so-called modes, that repeat every few years[133]. At a fundamental level, the existence of such modes resembles the recurrence of persistent spatial and temporal patterns exhibited by turbulent flows[134].

The dominant mode of climate variability at interannual time scales is the El Niño/ Southern Oscillation (ENSO), chosen here to exemplify ML contributions to interannual predictability. During an El Niño, which is the positive phase of ENSO, sea surface temperatures in the central and eastern Pacific Ocean increase by a few degrees with respect to average values; during the opposite phase of the oscillation, La Niña, the eastern Pacific is colder than average. Various indices that measure the area-averaged sea surface temperature anomalies (i.e. the deviation with respect to the mean seasonal cycle) in portions of the equatorial Pacific are used to characterize ENSO.

El Niño events, however, are difficult to predict due to their complexity which manifests through a spectrum of intensities, spatial patterns and temporal evolution[135]. Both statistical and dynamical models are routinely employed for ENSO prediction[136] with dynamical models generally outperforming the statistical ones. If initialized before boreal spring, models performance deteriorates compared to when they are initialized in summer. The latter notion, limiting skillful prediction beyond a 6 months lead time, has long been known as spring predictability barrier[135].

Much progress has been achieved in ENSO prediction using ML methods[6] and it is now clear that ML models can break the spring predictability barrier. CNNs trained on both model (using transfer learning) and reanalysis data have higher forecasting skill than most dynamical models. Their skill remains high up to 17 months lead time[137]. CNN-based schemes can also anticipate which type of El Niño pattern[135] will develop, and relatively shallow NN models using a relatively small predictor set have delivered skillful long-lead time predictions and quantified their uncertainty[138].

Recent research has focused on improving prediction of ENSO complexity for long lead times (> 9 months). Extension of the CNN framework[137] with different loss functions and heterogeneous parameters has achieved skillful forecasts up to 24 months[139], while Reservoir Computing methods (see Methods) have reached 21 months[140]. An all-season CNN model successfully improved forecast skill in boreal spring[141].

Lastly, XAI techniques (e.g. contribution maps) used together with CNNs have allowed to link ENSO precursors and physical processes[142], while graph neural networks, such as adaptive graph CNNs[143], have shown good prediction skill (up to 18 months) and improved explainability by learning (global) spatial patterns from data.

### 2.3.4 Decadal forecasting

Decadal prediction focuses on forecasting near-term climate change, on a scale of about ten years. At the intersection between an initial value and a boundary value problem, climate at decadal time scales is strongly modulated by ocean processes. Operational decadal prediction is coordinated by the World Meteorological Organization and is routinely achieved using a large ensemble of simulations from different climate models[144]. These models have persistent biases, insufficient resolution to capture baroclinic instability processes in the ocean and face large uncertainties in their initial conditions, especially in the ocean state[145]. Efforts to quantify the benefit of using models that explicitly resolve ocean vortices and the instabilities that generate them down to scales of about 10 km[146,147] are ongoing, for example within the EU-funded DYAMOND, DestinE and nextGEMS projects.

Fully data-driven ML approaches, similar to those introduced for weather time scales are prone to instability when used at long (many months) lead times. Grid distortions were identified as base motive for the instabilities, and the problem has been addressed in part by introducing Fourier neural operators, obtaining roll-outs of up to a year[148]. Recently, the time span and achieved stable predictions were extended to a decade[149], with mass conservation holding to a good approximation without being explicitly enforced. In addition, a spectral bias[150] was found to limit long term predictions, because training tunes NNs to the low-frequency content of the signal[151]. A multi-pronged approach that explicitly controls the high-frequency component of the prediction in the loss function and a self-supervised spectrum correction strategy[151] addressed the issue. Although observations below the ocean surface remain sparse, ocean emulators are emerging[152,153].

Online learning algorithms, of the type discussed in 2.3.2, which track an ensemble of predictors (in this case a climate model ensemble[154]), and foundation models, such as ClimaX[155], which are trained on global climate model outputs, have also shown skill on longer time scale predictions. Other advances include the application of NN to Earth system model output allowing, through XAI techniques, the identification of ocean surface temperature patterns that lend predictability on surface temperatures across North America[156].

Finally, hybrid architectures, such as NeuralGCM[74], are been explored to obtain multi-decadal projections. Within the Earth Visualization Engines (EVE) initiative[157], it is proposed to adopt ML techniques to improve training (AI-inside) and to learn simultaneously from model outputs and auxiliary data (AI on top). The latter approach, also used in DestinE, goes beyond emulation and is poised to create new types of (hybrid) models.

## 3 Current Challenges and Future Directions

Machine learning has enabled tackling a broad range of climate science problems in novel ways. These developments will likely accelerate in the coming years, influenced by the rapid progress of ML techniques and by the increasing corporate and philanthropic investments in climate science. So far, ML has broken long standing predictability barriers in climate prediction, has given new impetus to the discovery of equations governing components of the climate systems, and is revolutionizing weather forecasting through hybrid models or emulators as good or better than traditional models at a fraction of the compute cost.

The climate system, however, presents peculiar challenges that stems from its boundary value problem nature and its complexity. We conclude this review discussing what those challenges entail, and the most promising advances anticipated in the near future.

A key challenge of ML for any application is the lack of well constrained data. For the climate system, there exist less than 50 years of dense observational data (satellite era) that are strongly biased towards the atmosphere and the ocean surface, are spatio-temporally correlated and belong to a single realisation. For the reconstruction and assimilation applications discussed in Sec. 2.1, the data scarcity limits the generalization capabilities of the ML models, but local, specific reconstruction efforts have profited greatly.

For modeling the climate system, data availability limits both traditional and ML approaches. Indeed, conventional models rely on observational data for tuning parameters and parameterizations, as well as verification. In light of this commonality, we support the verification of ML-based approaches to be fundamentally the same as for conventional models: next to comparisons with the single observed climate realisation, other diagnostics that account for conservation laws and physical consistencies must be used[74,120,149]. The use of metrics that account for physical principles would mirror the development of ML-based weather prediction emulators[112,113].

To address the lack of data, several complementary approaches hold much promise. The first, just mentioned, is to develop hybrid models where conventional components are retained and either improved by, or complemented with, ML-based components; the data-driven parameterizations discussed in 2.2 are an example. The second is to develop emulators that at least partly rely on climate model simulations, as discussed in Sec. 2.3. Central to both is the question of how much physical knowledge should be incorporated into the ML algorithms. Work on emulators like ACE[149] and on data-based weather prediction[111,120] indicate that intrinsic physical properties can be learned when sufficient data is available. If this applies to most physical properties in the large data regime, and whether physical constraints can improve results[64] is unclear but an exciting area of fundamental research. This is especially important because most symmetries and conversation properties have been verified only for idealized realizations of the climate system.

A third novel way to confront data scarcity consists in developing holistic ML models that combine as much data as possible, e.g. global and local reanalyses, simulations and observations, with the various data sources complementing and correcting each other. Methodologically, this approach falls into the realm of large-scale representation learning[158] and is a foundation model[159] for weather that can be extended to climate. It can provide a task independent representation of the climate system that could be adapted for a range of tasks, from forecasting to process understanding. First steps in this direction can be found in[155,160], and those efforts are likely to multiply in the next few years. A major challenge remains the storage of all the model simulations needed at the temporal and spatial resolution required by ML to be most useful. A fourth direction leverages and combines common ML approaches for dealing with data imbalance (e.g., re-sampling, weighted loss functions)[53,129,161] and novel mathematical frameworks for rare-event sampling[162].

There is significant interest in developing climate emulators trained on the output of conventional climate simulations, building on the rapid progress in numerical weather prediction[94,103]. ML emulators can already deliver stable prediction of specific variables[148,149], at a fraction of the computational cost, although no model of the entire system yet exists. The potential offered by emulators in climate science is multi-fold. First, they would enable much larger ensembles than currently possible, being orders of magnitude faster. Second, they could interpolate between different scenario simulations or between simulations performed with different models. Third, emulators can compress the model output and distribute results much more efficiently, allowing for reconstructing state information at a higher spatial and temporal resolution than typically stored for conventional climate model simulations. This compression is critical if extremely high-resolution simulations[157] become reality, for seamless weather-to-climate prediction systems, or when climate data are needed for specific societal decision-making, for example for evaluating risks and economic impacts of the fast changing energy infrastructure, climate feedback associated with carbon dioxide removal technologies deployed at scale, or regional changes in extreme events.

Much effort has been placed on developing hybrid dynamical-ML models (see Secs. 2.2 and 2.3). While conventional model components ensure generalization capabilities and physical consistency, ML can reduce model biases and obtain representation of uncertainties and long-tail distributions inexpensively next to the traditional approach of increasing model resolution. The use of the conventional model components also alleviates the training data concern. As far as parametrization development, we expect continuing progress with new algorithms that focus on individual processes[55], but especially with new methodologies that represent entire climate sub-systems. A key challenge with complex systems is the representation of the interaction among highly nonlinear processes: ML models in principle can learn holistically an entire sub-system, with training being performed either on high-resolution, (quasi-)resolved model data or observational data, or both. However, successful hybrid models that show reduced biases against current climate simulations have been slow to emerge because of challenges in coupling NNs with climate models and accounting for interactions of several ML-based parameterizations. The new NeuralGCM[74] shows promise in this direction.

Climate projection systems entirely based on observations are likely to emerge as well. Machine learning will thereby contribute to extract new patterns in existing observations and we will learn about feedback and dynamics that are poorly represented in conventional models. A fully data-driven approach may be especially useful for producing reanalysis products which are observationally constrained state estimates of the climate system, but remain by and large determined, and limited by, forecasting models.

Skillfully trained ML models can contribute to discoveries and breakthroughs. Example include new insights on the relationships among slow and fast processes in the climate system, and the uncovering of physical behaviors, as happened when examining the improved ENSO forecast of ML models[163]. A better understanding of climate modes is indispensable to contextualize their evolution. Developing data-driven indices of such phenomena may answer outstanding questions on their past, present and future, while causal ML can point to relationships between distant anomaly patterns at different time-lags[164]. Decoding such information requires tools that harness recent developments in explainable AI[165,166]. In climate, as in many areas of physics, a key concern of the use of ML methods is indeed their lack of interpretability and explainability. We believe that coupling these advances with the adoption of existing, well tested, diagnostics and metrics will build trust[113,120].

Climate change is unequivocal, its potential consequences interest any societal realm. and accurate climate predictions are indispensable. We are in the early stage of a profound and rapid change in the field of climate physics, in which ML is poised to contribute to projections at much lower computational costs for operational use, more reliable and longer-term predictions, and improved model simulations with reduced biases and better representations of subgrid-scale processes.

# Methods: Machine Learning Techniques

Machine Learning is a broad field encompassing many methods, several of which have been applied to climate science. The list includes deep learning, probabilistic graphical models, causal analysis, online learning, clustering, and ensemble methods such as random forests. While this survey references multiple types of machine learning, given the breadth of the field, we will limit this section mostly to neural networks, the basis of the modern field of deep learning.

Neural networks can be understood as parameterized nonlinear mappings $N_\theta : V \to W$ between finite dimensional vector spaces $V$, $W$. By choosing a basis, the network can thus be seen as a map $N_\theta : R^n \to R^m$ (with the usual identifications for complex or other vector spaces). The parameters are denoted collectively as $\theta$, with $\theta \in R^K$, and they are determined to (approximately) optimally fit a relationship in observed data, a process known as neural network training.

**Neural network architectures** Neural networks are usually composed of a set of layers, i.e. $N_\theta = L_N \circ \cdots \circ L_1$, which simplifies their practical implementation and allows for efficient computations highly optimized for a small set of standard layers and layer components. The classical "perceptron" is a feed-forward layer of the form $L_i(x) = \sigma(W_i x + b_i)$ where $x \in R^{n_i}$, $W \in R^{n_i \times m_i}$ is the weight matrix, $b_i \in R^{m_i}$ the bias vector, and $\sigma(\cdot)$ is a nonlinear function applied element-wise to the affine transformation $W_i x + b_i$. Common examples for $\sigma(\cdot)$ are the sigmoid function, given by $e^x/(1 + e^x)$, for scalar $x$, or the ReLU function, which is zero for negative arguments and the linear function $x$ otherwise. It is often informative to consider $W_i$ as the adjacency matrix for a graph that "routes" the data through the network.

The composition of multiple layers of the above type is known as a multi-layer perceptron or a feed-forward neural network. The trainable network parameters $\theta$ are the entries of the weight matrices $W_i$ and bias vectors $b_i$. Another common neural network layer $L_i$ is a convolutional one, which is similar to a feed-forward layer but where $W_i$ is a circulant matrix that implements a (discrete) convolution. Convolutional layers substantially reduce the number of parameters compared to multi-layer perceptrons and they are useful for data on a regular grid with an approximate translation invariance, such as images. Another important layer type is self-attention. It takes as input a set or a sequence $\{x_i\}_{i=1}^r$ of so called tokens $x_1 \in \mathbb{R}^{n_i}$ and outputs an updated set or sequence $\{\tilde{x}_i\}_{i=1}^r$ given by

$$\tilde{x}_i = \sum_{j=1}^{r} \rho\left(x_i^T W_q^T W_k x_j\right)(W_v x_j) \tag{8}$$

where $W_q$, $W_k$, $W_v$ are projection matrices and $\rho$ is the *softmax* function, the generalization of the logistic sigmoid function (used in probabilistic, binary classification) to multiple classes. This can be thought of as smoothed version of the *argmax* that also ensures that the output is normalized and can be interpreted as a probability distribution. In essence, attention layers update each $x_i$ based on the dot product similarity to all other tokens and filtered by the softmax nonlinear function. A full attention layer usually consists of multiple parallel (and hence independent) so-called attention heads performing the computations in Eq. 8 with independent projections $W_q$, $W_k$, $W_v$.

Transformers, which are one of the most popular and successful neural network architectures to-date, consist of alternating attention and multi-layer perceptron layers. Conceptually, these update tokens with the inter-token attention computations in Eq. 8 and then map all tokens independently to a new vector space where the attention computation can be repeated, revealing different information. The layers are combined with so-called skip connections so that only a residual update is determined through the layers, i.e. $\bar{x}_j = x_j + L_i(x_j)$. Skip connections are typically used to improve the numerical stability of the training process. A variation of classical transformers is Fourier neural operators where the attention is computed in frequency space, exploiting the fact that a translation-invariant integral kernel can be expressed efficiently in the Fourier domain.

**Neural network training** Training of neural networks refers to the nonlinear optimization that is used to determine a set of optimal (or suitable) network parameters $\theta$ given a set of data $D$. In the simplest case, $D$ consists of pairs $D = (x_i, y_i)_i^D$ and the $y_i$ are the target quantities of interest that are to be estimated or predicted by the neural network. E.g. the $x_i$ can be images and the $y_i$ class labels that determine the image content. Directly training $x_i$ towards $y_i$ is known as supervised learning. When the $y_i$ are proxies to train the neural network but not themselves of interest then one speaks of self-supervised training. This is typically complemented with a second training phase towards a specific task or application. The purpose of self-supervised learning is to learn overall domain knowledge so that the model can represent the underlying structure and patterns of the data without relying on explicit labels. A classical example of self-supervised training is image inpainting, where parts of images are masked and the self-supervised training task of the network is to predict these parts.

The optimization that is required for training is typically solved with stochastic gradient descent where a Monte Carlo estimate of the gradient over a subset of the data is used in each gradient descent step. The Monte Carlo estimate is used since it is computationally cheaper and requires less computer memory. Empirically, it is also key to a successful optimization of neural networks with millions, billions or even trillions of trainable parameters since the stochasticity helps to avoid ineffective local minima. A wide range of energy or loss functions are used for the gradient descent. When the $y_i \in R^n$, then the mean squared error is, arguably, the most common loss function. When $y_i$ is from a discrete set, then cross-entropy is commonly used as loss.

A common problem with neural network training is *overfitting*, which is identified when the network performs well on the training set but not on new (unseen) data. To diagnose overfitting and enable an estimate of the performance of a trained neural network in applications, the data set is typically partitioned into training, validation, and test sets. Computing the skill of the neural network on the validation set is used to monitor training progress without affecting (and hence biasing) the network weights. In other words, the validation step helps with detecting overfitting or selecting *hyperparameters*, such as the number of layers and parameters in the neural network or the parameters of the stochastic gradient descent. The test set is employed only after training, to provide a final evaluation of the network performance in inference, i.e., when the trained network with fixed weights is put to its intended use (e.g, for prediction, estimation, or classification, etc.). Common strategies to address overfitting are to restrict the network size or use drop-out, which occurs when a random subset of the training weights is set to zero in each training optimization step. For sufficiently large training datasets, however, network size is no longer directly related to overfitting. Often very large (even over-parameterized) networks perform best[167], because parts of a large network can internally specialize for processing only specific inputs, for example a certain input variable at a specific height level in a weather model.

**Generative models** For classification problems, such as predicting the class of an image, network output is typically modeled not as a discrete value but as a probability distribution over all possible labels. This can be realized by predicting $y_i \in R^M$, where $M$ is the number of classes, and then normalizing the values to obtain a probability distribution, e.g. with the softmax function. The network can then be interpreted as modeling the conditional probability distribution $p(y|x)$ of outputs $y$ given the inputs $x$. The perspective of a neural network modeling a conditional or joint probability distribution is widely used in the machine learning literature and the corresponding models are referred to as generative models.

This principle can be extended to continuous probability functions $p$, with the network being trained to generate either moments of the probability distribution or samples from this distribution. For multi-dimensional data such as images or physical fields, generative adversarial networks (GANs) and diffusion models are common generative architectures with strong theoretical foundations that can be interpreted from a statistical mechanical point of view. In fact, any regression problem trained with mean squared error can be interpreted as a generative model where only the mean of a Gaussian distribution is considered.

**Incorporating physics into neural networks** Neural networks extract their power from the training data through the training process, with only weak priors through the network architecture (e.g. when a convolutional network is used). This stems largely from their development being driven by natural language processing and computer vision where no simple analytic theory is available. It is possible to incorporate prior physical knowledge following strategies explored more generally for physical systems[168]. Known symmetries[169,170], autocorrelations[171], and conservation laws[172] have been incorporated into neural networks. Physics-informed neural networks or PINNs replace the training set $D$ with a known constraint that the neural network output has to satisfy, typically a known physical equation[173].

**Reservoir Computing** A special class of methods are recurrent neural networks. In contrast to uni-directional feedforward neural networks, these networks are bi-directional, meaning that they allow the output from some nodes to affect subsequent input to the same nodes. Recurrent neural networks are difficult to train and several variant have been developed to circumvent that problem, such as Long-Short Term Memory networks and Reservoir Computing[174]. In the latter method, the input time series is mapped to a high-dimensional dynamical system (the reservoir) and only the output weights (connecting the reservoir nodes and the output) are adjusted during training. Reservoir Computing therefore has a close connection to dynamical systems theory and has been widely used in climate prediction studies[93,175].

**Explainability and interpretability** In recent years, there has been much research in machine learning on how to interpret the learned models. Some models are by nature interpretable (for example, linear regression yields a weighting over its input variables), but for others, especially deep neural networks, methods to provide *post hoc* "explanations" for model decisions have been developed. The AI subfield of Explainable AI (XAI) is concerned with understanding the reason why a neural network output was generated. *Post hoc* analysis of a deep neural network via an XAI technique can help assess the model's confidence in the decision, detect cases of inappropriate usage, and even derive physical insights from the neural network models themselves. In climate physics, where understanding is often more important than quantitative performance, XAI is a particularly active field [83,165,166]. We refer to[176] for a discussion of the challenges in choosing the best XAI method in the domain of climate science and for an extensive list of references.

## Open Research Section

No software/data is used in this paper.

## Acknowledgments

The authors thanks the Kavli Institute for Theoretical Physics and the University of California at Santa Barbara for their hospitality and excellent working facilities during November–December 2021. This research was supported in part by the National Science Foundation under Grant No. NSF PHY-1748958. The work of HD was funded by the European Research Council through the ERC-AdG project TAOC (project 101055096). PH was supported by the Office of Naval Research (N00014-20-1-2722). AB was supported by the U.S. Department of Energy (DE-SC0024709). JB was funded by the European Space Agency through the SuperIce project under the Contract No. 4000142335/231/I-DT. CM acknowledges the support of the French government's Choose France Chair in AI.